*Review Article*

# Atomic Cross-chain Swaps: Development, Trajectory and Potential of Non-monetary Digital Token Swap Facilities

**Mahdi H. Miraz[1,2,*] and David C. Donald[1]**

[1]The Chinese University of Hong Kong, Hong Kong SAR
m.miraz@cuhk.edu.hk; dcdonald@cuhk.edu.hk
[2]Wrexham Glyndŵr University, UK
m.miraz@glyndwr.ac.uk
*Correspondence: m.miraz@ieee.org

**Abstract:** Since the introduction of Bitcoin in 2008, many other cryptocurrencies have been introduced and gained popularity. Lack of interoperability and scalability amongst these cryptocurrencies was - and still is - acting as a significant impediment to the general adoption of cryptocurrencies and coloured tokens. Atomic Swaps – a smart exchange protocol for cryptocurrencies - is designed to facilitate a wallet-to-wallet transfer enabling direct trades amongst different cryptocurrencies. Since swaps between cryptocurrencies are still relatively unknown, this article will investigate the operation and market development thus far and query the advantages they offer and the future challenges they face. The paper contains detailed literature and technology reviews, followed by the main analysis and findings.

## 1. Introduction

In spite of the fact that many applications based on Blockchain [1,2], such as cryptocurrencies, including the pioneer Bitcoin, have proven to be disruptive, there are two major drawbacks inherent in the technology impeding application: limited interoperability between chains and limited scalability due to high transaction latency. Capped transaction throughput, resulting from the decentralised consensus approach and other network and algorithmic limitations, remains at the centre of all concerns associated with scalability of Blockchain based applications, especially for cryptocurrencies such as Bitcoin and Ethereum. To date, Bitcoin's transaction capacity is 7 per second on average while that of Ethereum is 15 per second; Ripple's capacity in this regard is much higher, 1500 transactions per second, compared to the others [3]. On the contrary, Visa can process 24,000 transactions per second on average. That being said, an individual "unconfirmed" transaction might take between 10 minutes to several days to get "confirmed" in Bitcoin network. The queue time in Ethereum is also exponentially increasing due to mushrooming amounts of ICO's (Initial Coin Offering), DAOs (Decentralised Autonomous Organization) and DApps (Decentralised Apps) running on the Ethereum network. Due to Blockchain's increasing popularity, high volume of R&D (Research and Development) is being conducted across the globe to overcome these scalability problems. In off-chain scaling [4,5,6,7,8,9,10], facilitating off-chain transactions (i.e. not registered on the Blockchain) to take place using two-party payment channels having two separate layers for transaction execution, interim transactions are updated off-chain while settled or netted on-chain. The Lightning Network (LN) [4,11]- implemented utilising HTLC (Hashed Timelock Contract) based smart contract- is the most technologically advanced two-layer off-chain solution.

While it is extremely important to scale up the Blockchain ecosystems to reduce transaction queue, lack of interoperability - especially for cross-chain swap of cryptocurrencies and other coloured coins such as deeds, stocks, bonds and intellectual properties - is another key problem that needs to be addressed to achieve the potential benefit from this technology. At this moment, there is no easy way to exchange coloured coins or cryptocurrencies such as Bitcoin (BTC) for Ripple (XRP). Dependence on legacy exchanges is a necessity. Obscurity of the coin might add an extra layer of difficulties leading to use of multiple exchanges, i.e. via several intermediary cryptocurrencies before converting into the desired one and thus generating high transaction costs for conversion.

This poses various problems and associated risks. For example, assume Alice possesses some BTC and would like to swap them for XRP. If Bob has some XRP and would like to swap them for BTC, in a *safe and secure* cross-chain swap enabled ecosystem, they could just swap cryptos at an agreed rate, as done with an ordinary swap contract involving fiat currency. Due to lack of chain interoperability, however, the parties now have to approach to a legacy exchange – run by a third-party administrator, sell their current cryptos and buy the desired one . If both cryptos are not traded on the same exchange, the buy-sell process become even more complicated and expensive. Apart from this, there are other significant problems and risks associated with such transactions, such as the exchange's vulnerability to hack, illiquidity, impositions of regulatory bodies, dishonest exchange mismanagement. These problems could have been eliminated if interoperability - enabling cross-chain swaps - was not an issue amongst various blockchains. This is where Atomic Swaps comes into the picture. As the name implies, Atomic Swaps are peer-to-peer (P2P) exchanges of one crypto asset for another between two counterparties (peers) by direct and automatic interaction of two separate blockchains at the binary level of their basic coding without the use of centralised intermediaries such as exchange. They are mainly conceived as functioning between parties that actually need the currency rather than as between a protection provider and a protection buyer in a traditional currency swap contract.

While many other proposals to solve these problems are being researched and evaluated such as Fusion, WanChain and Padlocks's parachain, recent development of Atomic Swap and Lightning Network appear to present more effective solutions. Considering the fact that both of these approaches are relatively new and still under development, this paper presents a comprehensive technical review of Atomic Swap, its current state, challenges and possible future applications.

## 2. The Evolution of Disintermediated Cross-chain Swap

As previously discussed, one cryptocurrency or coloured coin cannot usually be directly traded as each of these tokens exists only in its native blockchain. Traverse between different chains is prevented by a lack of interoperability. If two traders decide to trade their different cryptos with each other, an active intermediary will be required. This intermediary is usually a legacy crypto exchange. In fact, more than one exchanges might be required, depending on the obscurity of the crypto asset as well as the limited types of cryptos supported by any exchange at a given point of time. This multiplies transaction fees, exchange rate risk and commissions. This arises from an inherent failure of blockchain technology, but it is hoped that Atomic swaps may provide a workable solution.

**2.1. P2PTradeX**

It took almost four years to conceive a working proof-of-concept disintermediated cross-chain swap - since 2008 when blockchain was first introduced (through Bitcoin) till Sergio Demian Lerner presented the leading draft of a peer-to-peer swap called "P2PTradeX" [12] in July 2012 in Bitcoin forum on bitcointalk.org platform. Figure 1 presents how this P2PTradeX, also known as wallet-to-wallet trade, takes place.

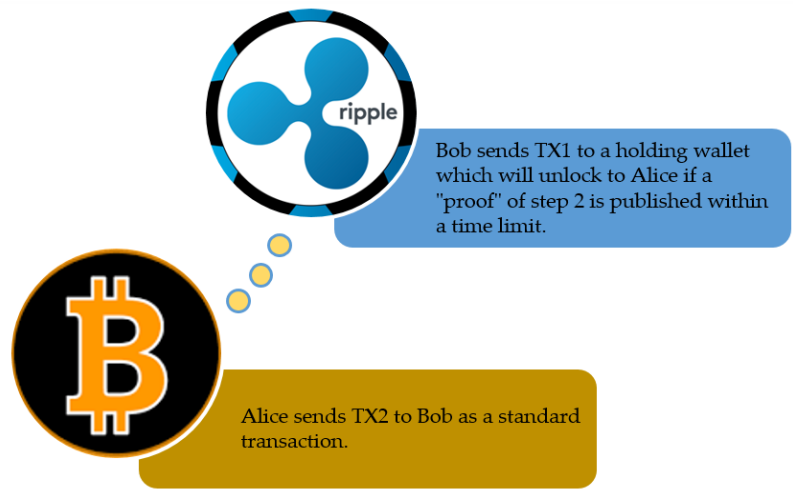

**Figure 1.** A simplified P2PTradeX Swap

While the swap may appear secure at a first glance, Alice in fact remains vulnerable to Bob's denial of the transaction. This is because her commitment is irreversible, while Bob's will depend on Alice establishing it, under a future proof burden lying with Alice. The notion presented in P2PTradeX was therefore incomplete.

**2.2. Atomic Cross-chain Swap**

The conceptualisation of "Atomic Swap" [13], as presented by Tier Nolan in May 2013 in the same Bitcoin Forum, made real improvement in achieving disintermediated cross-chain trade. This is considered to be the first full procedure for "atomic" transactions that occur at the binary core of the respective chains automatically, either entirely or not at all, and not subject to unwinding due to lack of acceptance.

Atomic swaps, also known as atomic cross-chain swaps or atomic cross-chain trades, are tête-à-tête cross-chain smart transactions which can arise between two nodes. They enable secure peer-to-peer interchange of two different cryptos without involving any broker or centralised intermediary, such as legacy crypto exchanges, for establishing enforceability. The term "atomic" has been borrowed from database systems terminology, where atomicity or an atomic transaction is limited to a set of binary outputs: guaranteed to occur either completely or not at all. Atomic swap thus eliminates the need for legacy exchange without the risk of one party defaulting on the trade. Atomic swap – being a cryptographically powered smart contract technology – enables peer-to-peer exchange of cryptos directly between two users while both of them having complete control and ownership of their old crypto until the transaction actually happens. Figure 2 illustrates a simplest form of such atomic swap.

After introduction of the concept, it took almost another four years to materialise the atomic swap. The first known effective execution of an atomic swap was between Decred (DCR) and Litecoin (LTC) on 20th September 2017 [14], followed by another successful swap between LTC and BTC three days later. Several other atomic cross-chain swap took place amongst various forks of Bitcoin or other cryptocurrencies based on Bitcoin protocol. Following their success and using the codebase of Decred, Altcoin.io Exchange also achieved swaps between BTC and Ethereum ERC-20 token [0.12345 ETH for 0.12345 BTC] [15] on 11th October 2017. Since then, several other decentralised exchanges, cryptocurrencies and startups such as 0x and Lightning Labs have incorporated atomic swaps to some extent. The latest addition in this race is the Lightning mainnet beta release (lnd 0.4-beta), by Lightning Labs, enabling instant off-chain atomic swaps between BTC and LTC, utilising bitcoin's lightning network.

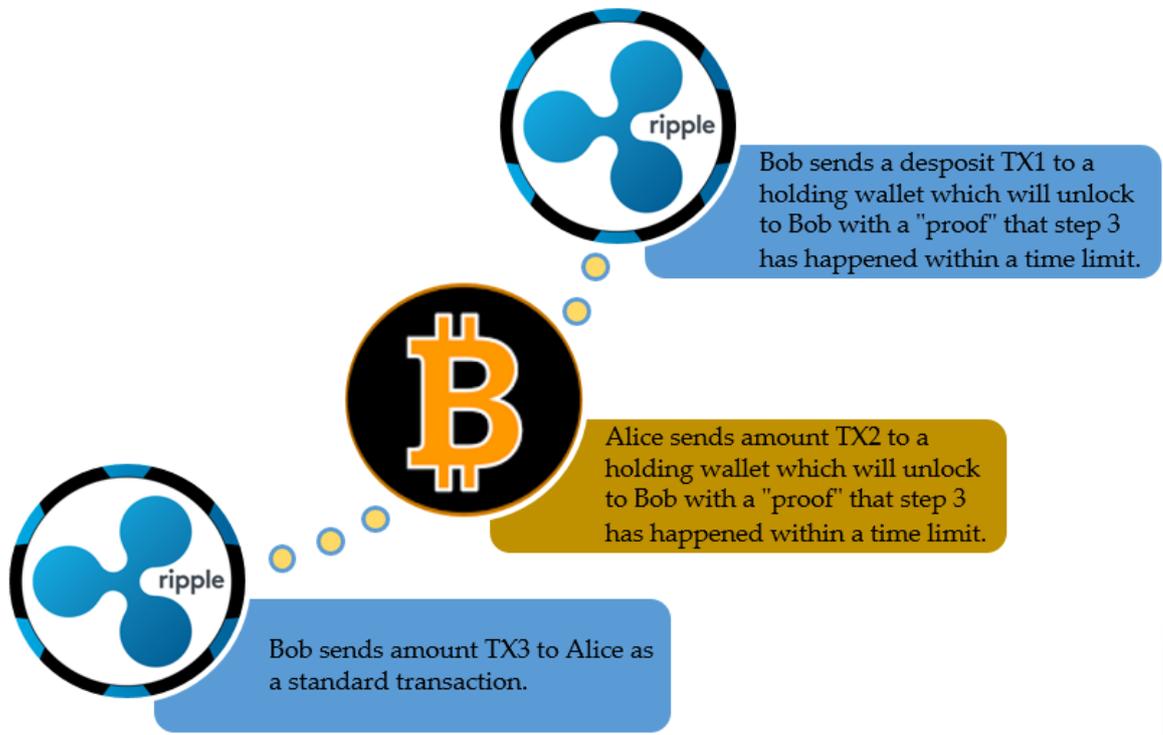

**Figure 2.** A simplified Atomic Cross-chain Swap

In fact, circa one year after Tier Nolan, another proof of concept was introduced on 18th April 2014 by an anonymous developer known as both "Jl777" and "James". He later became the principal developer of Komodo and led the development process of BarterDEX [16]. BarterDex cross-chain swap process included a few other provisions found in modern swap markets, such as order matching, clearing and settlement services, liquidity providers and time value premium. BarterDEX implements incentives for each completed stage of the swap and disincentives for deviation. Spam-deterrent fess as a protection against Denial of Service (DoS) attack and a security deposit which is 12.5% larger i.e. 112.5% of the amount to be traded. In addition, reputation scores for both counterparties are also recorded.

**2.3 Variants of Atomic Swaps**

Depending on where the interim transactions are being conducted, atomic swaps can be of two major types:
1. On-chain Atomic Swap
2. Off-chain Atomic Swap.

**2.3.1. On-chain Atomic Swap**

When an atomic cross-chain swap takes place between two different but homogeneous blockchain ecosystems, this is known as an on-chain atomic swap as the swap takes place directly on both the Blockchains.   The swaps between DCR and LTC as well as LTC and BTC as stated before are examples of on-chain atomic swap. A detailed description of how on-chain atomic swaps can be successfully implemented is presented in section 4.

**2.3.2. Off-chain Atomic Swap**

Off-chain atomic swaps take place, as the name suggests, on a separate layer (also known as "layer 2" or "second layer") away from the chains, such as Lightning Network of Bitcoin. Off-chain swaps support trading between both homogeneous and heterogeneous blockchain ecosystems. The atomic swap between BTC and Ethereum ERC-20 token by Altcoin.io Exchange is an example of a heterogeneous off-chain swap while the swap between BTC and LTC by Lightning Labs is an example of homogeneous off-chain atomic swap. Because of the use of Lightening Network, the swaps are instantaneous.

In its simplest form, two merchants agrees on a shared private key for off-chain swap of their cryptos and the swap will take place in a special channel if and only if their keys match. Because the key is not known to others, it is not possible to get forfeited. If we consider HTLC, as described in section 4.2, as linking of two blockchains together, the lightning network then is similar to linking two payment channels together enabling the required environment for off-chain an atomic swap.

**4. Technical Fundamentals of Atomic Swaps**

**4.1. The Procedure**

Let us consider previous example of Alice and Bob again. Instead of approaching a legacy exchange, the counterparties may decide to exchange their respective cryptocurrencies directly at an agreed rate. Absent atomic swaps, a minimum of two separate transactions on two different blockchains are required for this swap to take place: 1) transfer of Alice's BTC to Bob on Bitcoin network and 2) transfer of Bob's XRP to Alice on Ripple network. The consequences of this arrangement is that after completion of the first transfer, which cannot be unwound, the transferee will not lose his benefit by failing to perform the second transfer. By contrast, an atomic swap would take the following steps:

> Step-1: Alice creates a contract address (similar to a multi-lock safe)
> Step-2: Alice then generates a secret value and calculates the hash of it.
> Step-3: Alice then deposits her BTC to the contract address. [This BTC can only be claimed by fulfilling either of the two conditions: providing the value that generated the hash in step-2 signed by Bob's private key or signed by Alice's private key after a certain period of time elapses]
> Step-4: She sends the hash of the value and the address to Bob
> Step-5: Like Alice (in step-1), Bob generates a contract address but using the hash provided by Alice.
> Step-6: Similar to step-3, Bob sends his XRP to this contract address (of step 5). [This XRP can only be claimed by fulfilling either of the two conditions: providing the value that generated the hash in step-2 signed by Alice's private key or signed by Bob's private key after a certain period of time elapses. In the first case, value is automatically released to the network]
> Step-7: Alice gets her new XRP from this address using the secret value that generated the hash along with her signature and by revealing the value to Bob
> Step-8: Using the value received from Alice and Bob's private key, Bob gets his new BTC.

The intermediary would also be bonded in this arrangement. Despite the secret value is revealed to the network at time of Alice claiming her new coins, the intermediary could not steal Bob's coins since it also needs Bob's private keys for claiming them. The transactions are also time delayed so that if Bob does not transfer his coin to Alice or Alice changes her mind before Bob sends her the coins, both of them can claim their coins back, after certain amount of time has elapsed. Furthermore, Alice must reveal the secret value to the network to claim Bob's coin, as a result she cannot renege on the transaction. Thus, either both transactions required for the swap are fully consummated or neither. The following algorithm, as proposed by Tier Nolan [13], makes this atomic swap happen.

**Algorithm 1.** Atomic Swap Algorithm with Time Lock (adapted from [13])

**Set-up Phase**
    Alice chooses a random number, x and calculates hash, H(x).
    Alice generates transaction, TX1: "Transfer n1 BTC to <Bob's public key> if (x for H(x) known and signed by Bob) or (signed by Alice & Bob)"
    Alice creates transaction, TX2: "Transfer n1 BTC from TX1 to <Alice's public key>, locked 48 hours in the future, signed by Alice"    //This is the refund transaction timelocked by 48 hours.
    Alice sends TX2 to Bob
    Bob signs TX2 and returns to Alice    //Since now nothing has been publicly broadcasted to the network, therefore, nothing actually happens.

**Commitment Phase**
    Alice broadcasts TX1 to the network    // Alice can use refund transaction after 48 hours to get her money back as specified in TX2
    Bob creates TX3: "Transfer n2 XRP to <Alice's public-key> if (x for H(x) known and signed by A) or (signed by A & B)"
    Bob creates TX4: "Transfer n2 XRP from TX3 to <Bob's public key>, locked 24 hours in the future, signed by Bob"
    Bob sends TX4 to Alice
    Alice signs TX4 and sends back to Bob
    Bob submits TX3 to the network    // Bob can use refund transaction after 24 hours to get his money back as specified in TX2

**Claim Phase**
    Alice spends TX3 giving x
    Bob spends TX1 using x

Since the claim phase completes both transactions, Alice must spend her new coins within 24 hours; otherwise Bob can claim the refund and keep his coins. Similarly, Bob must spend his new coins within 48 hours; otherwise Alice can claim the refund and keep her coins. However, to be on the safe side, both of them should have completed the process well before the deadlines. Along with cryptographic hash functions, the process is also timelocked, therefore, it can be reversed regardless at what stage it is halted. The worst that can happen in this atomic swap is that the coins are locked for a certain amount of time if either party delays in processing the required steps or halts in the middle of processing. Further research needs to be conducted to refine this algorithm to eliminate this time delay problem. Use of lightening network could be one solution. However, at this moment only few crypto networks fulfil the minimum requirements for supporting lightning networks.

### 4.2. Hashed Timelock Contract (HTLC)

The process of combing both a cryptographic hash function as well as an imposed time delay through a smart contact is known as Hashed Timelock Contract (HTLC) as set out above in Algorithm 1. The HTLC requires successfully generation of a hash, which can be verified between the involved parties or by the network. In an HTLC powered atomic swap, both parties must acknowledge the receipt of the funds within a certain period of time using a cryptographic hash function. Failure to do so by both or either of the parties voids the entire trade, so that nothing is actually exchanged. Atomic swap thus eliminates counterparty risks ensuring the whole process is automatic.

### 4.3. Minimum Requirements

For successful completion of on-chain atomic swap, powered by HTLC, fulfilment of the following three conditions is a must:
1. Both blockchain ecosystems need to support same type of hashing function.
2. Both systems must support time locked contracts.

3. Both must support specialised programming functions to codify the swap algorithm.

In Bitcoin blockchain, these specialised programming functions are more commonly known as layered solutions: SegWit is the first layer while Lightning Network is the second layer – both help addressing the scaling problems.

**4.4. Major Advantages of Atomic Swaps**

There are many advantages that atomic cross-chain swaps can bring; the major ones are as follows:
- Because of increased interoperability, crypto currencies can better compete with fiat currencies.
- Legacy cryptocurrency exchanges have proven to be prone to attacks. Atomic swaps eliminate the need for such intermediation.
- Commissions and other associated fees charged by the legacy exchanges are eliminated, likely reducing total transaction costs.
- Atomic swaps, especially if off-chain, are mostly instantaneous.

**4.5. Major Drawbacks**

Since the concept of atomic cross-chain swap is in its infancy, it has yet to overcome many obstacles before it can be effectually adopted. The major drawbacks, at the time of writing this article include:
- Atomic swaps, especially on-chain ones, are very slow.
- If an agreed trade does not take place, it is reversible, but the refund is timelocked.
- Atomic swaps are not still supported in major wallets or exchanges.
- Crypto systems without smart contract support cannot facilitate atomic swaps.
- While atomic swaps solve part of the exchange problem, the need for fully decentralised exchange is not met. The concept of decentralised exchanges is to replace an exchange by a network – thus eliminating SPF (Single Point of Failure).
- Implementing atomic swap requires extensive programming skill.

**5. Multifaceted Applications and Future Trends**

There are many crypto systems which cannot technologically support atomic swaps at this moment, due mainly to lacking either HTLC or specialised programming functions. However, they may be able to embrace these features in the future.

The technology is in its infancy at this moment and the number of atomic swaps taking place is extremely small. However, such swaps are someday likely to become a fluid "background process". Interoperability between blockchains will increase the liquidity of the currencies arising in these blockchains and atomic swaps could serve as a link to take assets formulated in one chain into assets formulated in another, multiplying versatility without losing the smart function of the blockchain. This could include securities settlement systems. This would make not only the trades of cryptocurrencies and other assets more global, but extend the reach of the disintermediated transaction model represented by the blockchain.

From the perspective of private law, the governance of such swaps would be no more difficult than what is currently done through framework agreements governing international swap transactions. Choice of law and forum, as well as interpretation and default provisions could be written into the respective coding. From a regulatory perspective, however, atomic swaps could cross borders and legal systems, making them more difficult to monitor and regulate by the governments and their supervisory agencies. In off-chain atomic swaps, the transactions take places in a private channel while the netted result is broadcasted to the network of miners. Thus, apart from the participating merchants, no one actually knows the transaction details. This can lead to the growth of illicit markets. Although legacy exchanges are not fully regulated by governments, eliminating the legacy exchanges by atomic swap, either off-chain or on-chain, would take away the central focus of regulatory oversight and could make regulation even harder (unless inspection of chain coding is found to be sufficient). The future of cryptocurrency and thus atomic swaps may well hang from how legal and regulatory provisions are adjusted around the globe.

The technical problems associated with on-chain atomic swaps, especially transaction latency and concerns arising from time-lock, are likely to be addressed by lightning atomic swaps in the near future. Thus, parallel to the widespread adoption of blockchain technologies in various other domains [17,18], application of atomic swap will likely reach wider scope, especially in settlement and clearing of securities [19] and other coloured coins. This could also play an important role in bringing the direct holding of securities back to organized markets [20], providing higher transparency and better corporate governance. In such a direct system, securities holders could even be empowered to trade securities directly without the need for broker-dealers or agents. Another important aspect is enabling cross-listing facilities for blockchain based securities exchanges. Current models for cross listing of securities lets one company list its securities on more than one exchange [21]; atomic swaps could allow a similar function to take place via a smart protocol in the respective blockchains of future securities settlement systems.

As the atomic swap technology matures, the realistic options for application will become more concrete, particularly with regard to transfer and fungibility of other digital assets, such as tokens, data, securities or licenses.

### 6. Concluding Remarks

This paper explains the inherent limitations of interoperability and scalability of blockchain technologies and how atomic swap, powered by HTLC and Lightning Network, can solve these problems to some extent. The paper provides a detailed explanation of how the atomic swap works, along with its current state and future trends as well as possible applications, especially in the domain of coloured coins such as securities settlement.

### References

[1]   Mahdi H. Miraz, "Blockchain: Technology Fundamentals of the Trust Machine", Machine Lawyering, December 2017. Available: http://dx.doi.org//10.13140/RG.2.2.22541.64480/2

[2]   Mahdi H. Miraz and Maaruf Ali, "Blockchain Enabled Enhanced IoT Ecosystem Security", in proceedings of the International Conference on Emerging Technologies in Computing 2018 (iCETiC '18), Part of the Lecture Notes of the Institute for Computer Sciences, Social Informatics and Telecommunications Engineering (LNICST), vol. 200, London, UK, 2018, pp. 38-46. Available: https://link.springer.com/chapter/10.1007/978-3-319-95450-9_3

[3]   Richard MacManus, "Blockchain speeds & the scalability debate", Blocksplain, February 2018. Available: https://blocksplain.com/2018/02/28/transaction-speeds/

[4]   Joseph Poon and Thaddeus Dryja, "The Bitcoin Lightning Network: Scalable Off-Chain Instant Payments", Lightning Network, White Paper DRAFT Version 0.5.9.2, January 14, 14 January 2016. Available: https://lightning.network/lightning-network-paper.pdf

[5]   Pavel Prihodko, Slava Zhigulin, Mykola Sahno, Aleksei Ostrovskiy, and Osuntokun Olaoluwa, "Flare: An Approach to Routing in Lightning Network", Bitfury White Paper, pp. 1-40, July 2016. Available: https://bitfury.com/content/downloads/whitepaper_flare_an_approach_to_routing_in_lightning_network_7_7_2016.pdf

[6]   Christian Decker and Roger Wattenhofer, "A Fast and Scalable Payment Network with Bitcoin Duplex Micropayment Channels", in Proceedings of the International Symposium on Stabilization, Safety, and Security of Distributed Systems (SSS 2015), Part of Lecture Notes in Computer Science (LNCS), vol. 9212, Cham, Switzerland, 2015, pp. 3-18. Available: https://link.springer.com/chapter/10.1007/978-3-319-21741-3_1

[7]   Andrew Miller, Iddo Bentov, Ranjit Kumaresan, Christopher Cordi, and Patrick McCorry, "Sprites and State Channels: Payment Networks that Go Faster than Lightning", November 2017. Available: https://arxiv.org/pdf/1702.05812

[8] The Raiden Network. Raiden Network. Available: https://raiden.network

[9] Conrad Burchert, Christian Decker, and Roger Wattenhofer, "Scalable Funding of Bitcoin Micropayment Channel Networks", in Proceedings of the International Symposium on Stabilization, Safety, and Security of Distributed Systems (SSS 2017), Part of Lecture Notes in Computer Science (LNCS), vol. 10616, Cham, Switzerland, 2017, pp. 361-377. Available: https://link.springer.com/chapter/10.1007/978-3-319-69084-1_26

[10] Rami Khalil and Arthur Gervais, "Revive: Rebalancing Off-Blockchain Payment Networks", in Proceedings of the 2017 ACM SIGSAC Conference on Computer and Communications Security (CCS '17), Dallas, Texas, USA, 2017, pp. 439-453. Available: https://dl.acm.org/citation.cfm?id=3134033

[11] Marco Conoscenti, Antonio Vetrò, Juan Carlos De Martin, and Federico Spini, "The CLoTH Simulator for HTLC Payment Networks with Introductory Lightning Network Performance Results", Information, vol. 9, no. 9, September 2018, 223. Available: https://www.mdpi.com/2078-2489/9/9/223

[12] Sergio Demian Lerner, "P2PTradeX: P2P Trading between cryptocurrencies", Bitcoin Forum, July 2012. Available: https://bitcointalk.org/index.php?topic=91843.0

[13] Tier Nolan, "Alt chains and atomic transfers", Bitcoin Forum, May 2013. Available: https://bitcointalk.org/index.php?topic=193281.0

[14] Jake Yocom-Piatt, "On-Chain Atomic Swaps", Decred Blog , September 2017. Available: https://blog.decred.org/2017/09/20/On-Chain-Atomic-Swaps/

[15] Altcoin.io Exchange, "The First Ethereum <>Bitcoin Atomic Swap", Altcoin.io Exchange, October 2017. Available: https://blog.altcoin.io/the-first-ethereum-bitcoin-atomic-swap-79befb8373a8

[16] Jl777, "BarterDEX – A Practical Native DEX", BarterDEX Whitepaper, November 2016. Available: https://github.com/SuperNETorg/komodo/wiki/BarterDEX-%E2%80%93-A-Practical-Native-DEX

[17] Mahdi H. Miraz and Maaruf Ali, "Applications of Blockchain Technology beyond Cryptocurrency", Annals of Emerging Technologies in Computing (AETiC), vol. 2, no. 1, pp. 1-6, January 2018. Available: http://aetic.theiaer.org/archive/v2n1/p1.pdf

[18] Md Mehedi Hassan Onik, Mahdi H. Miraz, and Chul-Soo Kim, "A Recruitment and Human Resource Management Technique Using Blockchain Technology for Industry 4.0", in Proceeding of Smart Cities Symposium (SCS-2018), Manama, Bahrain, 2018, pp. 11-16.

[19] Mahdi H. Miraz and David C. Donald, "Application of Blockchain in Booking and Registration Systems of Securities Exchanges", in proceedings of the IEEE International Conference on Computing, Electronics & Communications Engineering 2018 (IEEE iCCECE '18), Southend, United Kingdom, 16-17 August 2018. Available: https://arxiv.org/pdf/1806.09687.

[20] David C. Donald and Mahdi H. Miraz, "Can distributed ledger technology return securities settlement to direct holdings?", Manuscript on file with authors.

[21] David C. Donald, "Networked Securities Markets: From Cross-Listing to Direct Connection" in The Research Handbook on Asian Financial Law, eds. Doug Arner, Andrew Godwin, Wan Wai Yee and Shen Wei (Edward Elgar, forthcoming 2019)